# Synthesis, characterization and computational simulation of graphene nanoplatelets stabilized in poly(styrene sulfonate) sodium salt

Celina M. Miyazaki[a,b], Marco A. E. Maria[a,b], Daiane Damasceno Borges[c], Cristiano F. Woellner[c], Gustavo Brunetto[c], Alexandre F. Fonseca[c], Carlos J. L. Constantino[d], Marcelo A. Pereira-da-Silva[e,f], Abner de Siervo[c], Douglas S. Galvao[c], Antonio Riul Jr[c,*]

[a] UNESP – Univ Estadual Paulista, Bauru, SP, Brazil

[b] Universidade Federal de São Carlos – DFQM, Sorocaba, SP, Brazil

[c] Instituto de Física *GlebWataghin* – DFA – UNICAMP, Campinas, SP, Brazil

[d] UNESP – Univ Estadual Paulista, Presidente Prudente, SP, Brazil

[e] Centro Universitário Central Paulista – UNICEP – São Carlos, SP, Brazil

[f] Instituto de Física de São Carlos – USP - São Carlos, SP, Brazil

## Abstract

The production of large area interfaces and the use of scalable methods to build-up designed nanostructures generating advanced functional properties are of high interest for many materials science applications. Nevertheless, large area coverage remains a major problem for pristine graphene and here we present a hybrid, composite graphene-like material soluble in water, which can be exploited in many areas, such as energy storage, electrodes fabrication, selective membranes and biosensing. Graphene oxide (GO) was produced by the traditional Hummers´ method being further reduced in the presence of poly(styrene sulfonate) sodium salt (PSS), thus creating stable reduced graphene oxide (rGO) nanoplateles wrapped by PSS (GPSS).

* Corresponding Author: Tel: +55 19 3521-5336. E-mail address: *riul@ifi.unicamp.br* (Antonio Riul Jr)

Molecular dynamics simulations were carried out ot further clarify the interactions between PSS molecules and rGO nanoplatelets, with calculations supported by FTIR analysis. The intermolecular forces between rGO nanoplatelets and PSS lead to the formation of a hybrid material (GPSS) stabilized by van der Waals forces, allowing the fabrication of high quality layer-by-layer (LbL) films with polyalillamine hydrochloride (PAH). Raman and electrical characterizations corroborated the successful modifications in the electronic structures from GO to GPSS after the chemical treatment, resulting in (PAH/GPSS) LbL films four orders of magnitude more conductive than (PAH/GO).

## 1. Introduction

Graphene is a milestone in carbon-based and 2D soft-materials due to its unique mechanical, thermal, electrical and optical properties, which led to experimental and theoretical studies and applications in energy [1–3], sensing [4], devices [5,6] and the production of hybrid materials featuring new properties [7]. It is a one atom thick sheet made of $sp^2$ carbons arranged in a honeycomb lattice that was firstly obtained by mechanical exfoliation of pyrolytic graphite [8]. However, despite its remarkable properties and variety of applications it is not trivial the production of pristine graphene in large areas, even considering recent advances in chemical vapour deposition (CVD) methods [9] that require high temperatures (300 – 500ºC) and expensive equipment.

Chemical synthesis is an alternative approach enabling a moderate fabrication of graphene-based materials under mild conditions, with added benefits of functionalization that allows the formation of hybrids or composites with polymers, nanoparticles, DNA, etc. [10]. Usually, the chemical process begins with graphite exfoliation by the Hummers' method [11], generating graphene oxide (GO), an insulating material with functional oxygen groups on the basal planes and at the edges of the formed nanoplatelets. The oxidation process makes GO easily dissolved in water and it can be further processed to reduced graphene oxide (rGO), which is a more conductive material resembling the graphene properties. The highest observed conductivity in rGO is explained by the partial reestablishment of the carbon $sp^2$ network after the chemical reduction process [12,13], and despite being less conductive than pristine graphene rGO is an attractive material for interfacial applications [14].

The use of graphene-based materials usually requires the deposition on a solid substrate such as Si, quartz or glass. Within this context, the Layer-by-Layer (LbL) technique is an alternative and versatile way to promote graphene-functionalized surfaces as it allows for well-defined molecular level control over thickness, morphology and structure, designing advanced functional nanostructures in a diverse range of applications onto practically any surface [15]. LbL films of graphene derivatives have been widely used in energy applications [16–18], selective membranes [19,20], sensing and biosensing [4,21,22].

There is a plethora of studies with LbL films of graphene-based materials, such as composites with polymers used in electromagnetic interference shielding at GHz range [23]. Lee and co-workers [24] achieved transparent, conducting LbL films using chemically modified rGO with controlled thickness, and different three dimensional GO structures could be formed throughout a new diffusion-driven layer-by-layer assembly process [25]. Multi-walled carbon nanotubes and rGO composites were employed in point-of-care and clinical screening of multiple diseases [26], and the assembly of carbon nanomaterials with conducting polymers was used to buildup 3D nanoarchitectures for energy storage and sensing applications [27]. Since the LbL assembly is based on the spontaneous adsorption of material from aqueous suspensions [15,28,29], it is of great importance the development, synthesis and characterization of water soluble graphene-based materials that can be further studied in distinct applications.

In this work, we present the chemical synthesis and characterization of rGO nanoplatelets functionalized with sodium polystyrene sulfonate (PSS), a water soluble material called here as GPSS. Molecular dynamics (MD) simulations were carried out to provide further insights on the self-organization of GPSS structure. It forms stable suspensions in water for ~ 10 weeks and it is attractive for LbL assemblies; expanding the possibility of applications in energy conversion, barrier properties, sensing and biosensing by the fact that pristine rGO is poorly dispersed in several known solvents. Fourier transform infrared spectroscopy (FTIR) analysis helped to identify the role played by van der Waals interactions between PSS and rGO nanoplatelets in the formed GPSS nanostructure. Polyalillamine hydrochloride (PAH) was used as a positively charged polyelectrolyte to build up both (PAH/GO) and (PAH/GPSS) LbL nanostructures with controlled thickness and morphology. Measurements made by atomic force microscope (AFM) indicated the presence of larger GO nanoplatelets when compared to GPSS, and impedance characterization shows that the (PAH/GPSS) film is four orders of magnitude more conductive than (PAH/GO), thus confirming the efficacy of the reduction process in the GPSS formation.

## 2. Materials and Methods

### 2.1. Materials

Graphite powder (98%), $H_2SO_4$ (95%), $KMnO_4$(99%), $K_2S_2O_3$ (99%), $P_2O_5$, $H_2O_2$, $H_6N_2O_4S$, were of analytical grade and used in the synthesis processes. Poly(styrenesulfonate) sodium salt - PSS (Mw = ~ 70000) and poly(allylamine hydrochloride) - PAH (Mw ~ 15000) were purchased from Sigma-Aldrich and used as received. Ultrapure water (18.2 MΩ.cm) acquired from a Millipore Direct Q5 system (at 25 °C) was used in all experiments.

*2.2. GO and GPSS nanoplatelets synthesis*

A pre-oxidation process of graphite was performed in accordance to Kovtyukhova *et al.* [30] in which graphite (10 g) was immersed in a mixture of concentrated $H_2SO_4$ (15 mL), $K_2S_2O_8$ (5 g) and $P_2O_5$ (5 g) at 80°C, reacting for 6 hours at room temperature. Water was carefully added to dilution and the solid was washed till neutral pH. Sequentially, the Hummers´ method [11,30] was applied by adding 10 g of pre-oxidized graphite into concentrated $H_2SO_4$ (230 mL) (cold bath at 0°C). 30 g $KMnO_4$ were added slowly with constant stirring with extreme caution as the temperature should not exceed 20°C. The mixture was stirred for 2 h at 35°C. Water (460 mL) was added and allowed to react for 15 min, and then 1.4 L of water and 30% $H_2O_2$ (25 mL) were added to stop the reaction. The resulting yellow product was filtered, washed with 1:10 HCl solution and dried in vacuum oven at room temperature to produce GO.

For the GPSS synthesis, PSS and GO were added according to Stankovich *et al.* [31]. Briefly, 50 mL of GO suspension (1 g.L$^{-1}$) were prepared in ultrasonic bath for 30 minutes and mixed with 500 mg of PSS. The mixture was kept under vigorous stirring and hydrazine sulfate was added to a final concentration of 0.01 mol.L$^{-1}$. The final product (GPSS) was stirred and maintained at 90°C for 12h, being further washed with ultrapure water and dried at 90°C in vacuum oven. GPSS was characterized with UV-vis absorption spectroscopy using a Thermo Scientific Genesys 6 equipment, while Raman spectroscopy was performed with a micro-Raman Renishaw, model in-Via, laser at 633 nm. FTIR spectroscopy was acquired in a Thermo Nicolet, model Nexus 470, while X-Ray powder diffraction (XRD) analysis was performed in a Rigaku Rotaflex, model RU200B, Cu anode and λ = 1.542 Å. The chemical composition of the samples was analyzed by X-ray photoelectron spectroscopy (XPS) in a homemade setup using a vacuum chamber (10 mbar) and Mg Kα radiation (1253,6 eV) with 240 W. Shirley background was assumed to the XPS spectra fitting. Electrical measurements were acquired in a homemade four-probe setup in accordance with [32], using drop cast films of GO and GPSS deposited on quartz substrates, with film thickness measured by a Veeco Dektak 150 Surface Profilometer.

## 2.3. Molecular Dynamics Simulations

Molecular dynamics (MD) simulations were carried out to study the rGO nanoplatelets enveloped by PSS and water molecules. The preparation of the initial configuration of the system consisted in randomly placing the PSS and water molecules around one single rGO nanoplatelet. The PSS was built with 16 monomers of sulfonated styrene group (see Figure 1). Every sulfonate group is treated as fully ionized and $Na^+$ was added to neutralize the system. The water content is 5 water molecules per sulfonate groups. The rGO sheet has dimensions of ~ 40x40 Å (~ 1600 Å$^2$ in area), with the atomic concentration of the functional group been 2% hydroxyl, 2% epoxy, 1% carboxyl and 1% carbonyl. The functional group percentages are based on the XPS data discussed along this manuscript. The hydroxyl and epoxy groups are situated away from the edges, while carboxyl and carbonyl groups are on the boundaries of the nanoplatelets [53,58,59], as shown in Figure 1. Once the initial configuration was set, the system was equilibrated after a series of annealing and optimization runs using Berendsen thermostat and barostat [33]. After equilibration at ambient temperature and pressure (*i.e.* T=300 K and P=1 atm), canonical trajectories of 2 ns using Nosé-Hoover integrator scheme [34,35] were generated for analysis.

The pair-atom interactions were described within the classical formalism. The water molecules were described by rigid 3-sites Transferable Intermolecular Potential (TIP-3P) [36] model. To simulate the rGO nanoplatelets, all atoms were considered explicitly in the simulations (*all-atom* approach), whereas a mixed representation were used to describe the PSS molecule. The united atom model was used to describe the backbone chain, while the all atomistic representation for the benzene ring and sulfonate groups. The force field parameters, including bonded and non-bonded interactions for the PSS were extracted from Ref. [37], while for the rGO were extracted from Ref. [38]. The initial configuration was built using PACKMOL [39] and the MD simulations were carried out using LAMMPS [40].

*2.4. LbL films assembly and characterization*

LbL films were fabricated using PAH as positive polyelectrolyte (1 mg.mL$^{-1}$) for both GO and GPSS that were used as negative charged suspensions (0.1 mg.mL$^{-1}$), forming (PAH/GO)$_n$ and (PAH/GPSS)$_n$ films, being *n* the number of deposited bilayers. All solutions were adjusted to pH 3.5 with HCl solution and a washing was performed at each deposition step using ultrapure water also at pH 3.5 to remove material loosely bound. The immersion time was equal to 3 min. for PAH and 5 min. for GO and GPSS solutions. The LbL film fabrication was made using an automated dipper from a Langmuir trough (NIMA Technology, model 612D), with upward and downward rates of 60 mm.min$^{-1}$ and 10 mm.min$^{-1}$, respectively.

LbL films were characterized using UV-vis spectroscopy (Thermo Scientific Genesys 6), AFM (Bruker Dimension ICON Nanoscope-V) at intermittent contact mode with a silicon cantilever (40N/m, 330 kHz) and impedance measurements (Solartron impedance analyzer 1260A, coupled to a 1296 Dielectric Interface) acquired from 1 MHz −1 Hz, using 20 mV amplitude. (PAH/GO)$_5$ and (PAH/GPSS)$_5$ LbL films were deposited onto gold interdigitated electrodes (IDEs) having 30 pairs of digits, 3 mm long, 40 μm width and separated 40 μm each other. The IDEs were fabricated at the Brazilian Nanotechnology National Laboratory (LNNano/CNPEM - Campinas, Brazil).

### 3. Results and discussions

GO and GPSS aqueous dispersions were characterized by UV-vis absorption spectroscopy (Figure 2). The dispersions were stable for ~ 10 weeks, demonstrating that the rGO is indeed dispersed more effectively in presence of PSS. In Figure 2 GO solution displays bands at 229 and 282 nm, characteristic of π – π* transition of aromatic C–C bond, and n – π* transition of C=O bond, respectively [41,42]. GPSS solution presents a band at 268 nm due to π – π* transition (shifted to higher wavelength), suggesting that the electronic sp$^2$ conjugation was restored [41], and a band at 227 nm due to the PSS benzene group adsorption [43].

Figure 3 presents the Raman spectra of powder samples from graphite, GO, pristine rGO and GPSS. The characteristics G-band at ~ 1590 cm$^{-1}$ from the E$_{2g}$ degenerate mode zone vibration [44] of the sp$^2$ carbon network, and the D-band at ~ 1330 cm$^{-1}$ attributed to defects in the sp$^2$ carbon lattice [45,46] were investigated. D and G peak positions were only slightly shifted to higher wavenumbers (see Table 1) comparing pristine rGO with GPSS. This suggests that PSS is not chemically adsorbed on the rGO sheet, consequently not affecting the sp$^2$ network domains. The I$_D$/I$_G$ ratio increased after the GO reduction to form rGO and GPSS, suggesting a decreased average size of the sp$^2$ carbon domains [47,48], which means that new sp$^2$ domains were created, smaller in size and more numerous than those present in GO (before reduction).

The atomic ratio of C and O was obtained from XPS survey spectra (Figure S1), and Figure 4 shows XPS analysis of GO and GPSS, with C=C bonds increasing from 48.7% to 74.3% after the chemical reduction, and O/C ratio decreasing from 51.4% to 7%, indicating a successful reduction process [13]. Table 2 presents the relative percentage of each functional group, which is an important data for the computational analysis. The high-resolution C 1s spectrum of GO indicates high degree of oxidation with three main components: C–C from non-oxygenated ring (284.5 eV), C in C–O bonds (286.6 eV), and C in C=O from carbonyl groups (288.5 eV). The GPSS spectrum indicated a marked decrease in the oxygenated groups and an additional peak at ~285 eV related to C–S from PSS [49].

XRD analysis (Figure S2) showed graphite main peaks at $2\theta$ = 26.5°, 44.4° and 54.5° corresponding, respectively, to (0 0 2), (1 0 1) and (0 0 4) plans [50]. After oxidation, characteristic graphite peaks disappeared, rising a peak at $2\theta$ = 10° from (0 0 2) GO plan [50,51]. The interplanar distance ($d$) was shifted from 0.34 nm (graphite) to 0.90 nm after oxidation, which is attributed to the addition of oxygen groups and water molecules between layers [50], thus weakening van der Waals forces and assisting the exfoliation process. After the chemical reduction with hydrazine using PSS as stabilizing agent, thus forming GPSS, the GO peak ($2\theta$ = 10°) becomes extinct and a new broad peak appears at $2\theta$ ~ 20°.

FTIR analysis (Figure S3) indicated typical bands for GO at 1626 cm$^{-1}$ (C=C stretching), and the presence of oxy-groups at 1739 cm$^{-1}$ (C=O stretching), 1057 cm$^{-1}$ (C–O stretching), 1405 cm$^{-1}$ (C=OH stretching) and 1225 cm$^{-1}$ (epoxy group) [52], with a broad absorption band at ~3400 cm$^{-1}$ (C–OH stretching) [53]. After the reduction process with hydrazine and PSS, the oxy-group bands disappeared and comparing PSS with GPSS spectra the benzene ring vibration in PSS was upshifted from 1128 to 1132 cm$^{-1}$, indicative of physical interactions between PSS and the rGO, hampering the vibrations of the PSS chains. The shift in the FTIR spectra can be related with a strong interaction between the PSS molecules with the rGO sheets.

In order to better understand the GPSS structure, classical MD simulations were performed. Figure 1 shows a typical snapshot of the self-organized system containing PSS, water and rGO well thermalized at ambient temperature and pressure. The typical mass density is approximately 1.4g/cm$^3$, which is higher than 0.8 g/cm$^3$ PSS in bulk. Figure 5.a shows the radial distribution function between Oxygen atoms of the sulfonate groups and Hydrogen of OH and COOH functional groups. The simulations revel that the PSS interacts with rGO mainly through the hydroxyl and carboxyl groups forming H-bonds (see Figure 5). Moreover, the water molecules intermediate this interaction also via H-bonds as shown in RDF of water-functional group water-SO$_3$ presented in Figure 5.b. The PSS is found in stretched conformation with the sulfonated styrene groups "lying" on rGO (see Figure 1.b). This conformation maximizes the $SO_3^-$ - $OH$ interactions and minimize the electrostatic repulsion among sulfonate neighbors. This interaction should be strong enough to trap the styrene groups and, thus, hamper the vibration of PSS chain as suggested by FTIR analysis. These results are in good agreement with Raman analysis [54] indicating that there is a trend to PSS accommodation close to rGO nanoplatelet without chemical reaction.

After characterization of GPSS and GO composites, PAH polyelectrolyte was used to build up both (PAH/GO) and (PAH/GPSS) LbL nanostructures. LbL film growth was investigated using UV-vis absorption spectroscopy, illustrated in Figure 6. It was initially determined the best immersion time in each polyelectrolyte (results not shown), with both LbL architectures (PAH/GO)$_n$ and (PAH/GPSS)$_n$ displaying a linear rise in absorbance with increasing *n*. The observed linearity suggests that the same amount of material was adsorbed at each deposition step in the LbL film formation.

Figure 7 shows AFM analysis for (PAH/GO)$_5$ and (PAH/GPSS)$_5$ films. (PAH/GO)$_5$ presented platelets with lateral dimensions ranging from 10 to 30 µm, while it was observed smaller platelets in the (PAH/GPSS)$_5$ film, as already predicted by $I_D/I_G$ ratio by Raman spectroscopy, with lateral dimensions ranging from 1 to 5 µm. The height profile in Figure 7a displays a large fragment of a wrinkled GO platelet in the film surface (~ 50 nm), where apparently, all the background surface is covered with thin GO layers, and in Fig. 6b it can be seem smaller platelets for GPSS (~ 10 nm height). LbL film thickness analysis was performed with the AFM tip scratching (PAH/GO)$_5$ and (PAH/GPSS)$_5$ films (results not shown) indicated an average thickness of ~ 2 nm per deposited PAH/GO bilayer and ~ 4 nm per deposited PAH/GPSS bilayer, reinforcing the rGO wrapping by PSS molecules, which is compatible to the monolayer thickness reported in literature [55].

Functionalized, water soluble GPSS nanoplatelets were successfully formed presenting better electronic properties than GO. Figure 8A shows the impedance data obtained from 1Hz – 1MHz modulus for LbL (PAH/GO)$_5$ and (PAH/GPSS)$_5$ films. As expected, (PAH/GO)$_5$ presented a dielectric material profile, with higher impedance in the low frequency region (> $10^8$ Ω) followed by a capacitive behavior > 100 Hz (Fig. 8B). (PAH/GPSS)$_5$ LbL film displayed a more conductive behaviour (Fig. 8) with impedance values ~ four times lower than those presented by (PAH/GO)$_5$.

DC measurements acquired from the four-probe technique (see Figure S4) indicated conductivity values of 2.0 x $10^{-2}$ S.m$^{-1}$ for GO [56] and 6.3 x $10^2$ S.m$^{-1}$ for GPSS, similar to values found in the literature [41,47,57] and lower than those found for printed pristine graphene

($\sim 2.5 \times 10^4$ S.cm$^{-1}$) [58]. This difference can be explained by defects and residual oxygen containing groups introduced in the rGO nanoplatelets from the chemical reduction process [59], remarkedly impacting the electrical properties of GPSS when compared to pristine graphene. Nevertheless, our results corroborate the information of better electronic properties for GPSS when compared to GO [60] due to a better reestablishment of the sp$^2$ carbon network in the formed rGO nanoplatelets, which is not disturbed owing to the non-covalent interactions between PSS molecules, in close agreement with the MD simulation, Raman and FTIR analysis.

## 4. Conclusions

The present work demonstrated a successful synthesis to form GPSS nanoplateles that are easily soluble and form stable water suspensions. The presence of functional groups in GPSS favors physical interactions between PSS molecules with the rGO nanoplatelets, thus forming a composite hybrid material with interesting electrical properties. The atomic structuration of the materials forming the GPSS structure is essentially attributed to H-bond formation between oxygen groups in PSS with hydrogen from hydroxyl and carboxyl groups in rGO. Those interactions are favored even in the presence of water, becoming an interesting system to build up ordered multilayer nanostructures using the LbL assembly. (PAH/GO) and (PAH/GPSS) LbL films displayed a good linear growth, with the GO film presenting larger nanoplatelets when compared to GPSS due to the chemical attack during the chemical reduction in the GPSS formation. The non-covalent interactions between PSS and rGO endow interesting electrical properties to GPSS, which is four orders of magnitude more conductive than GO.

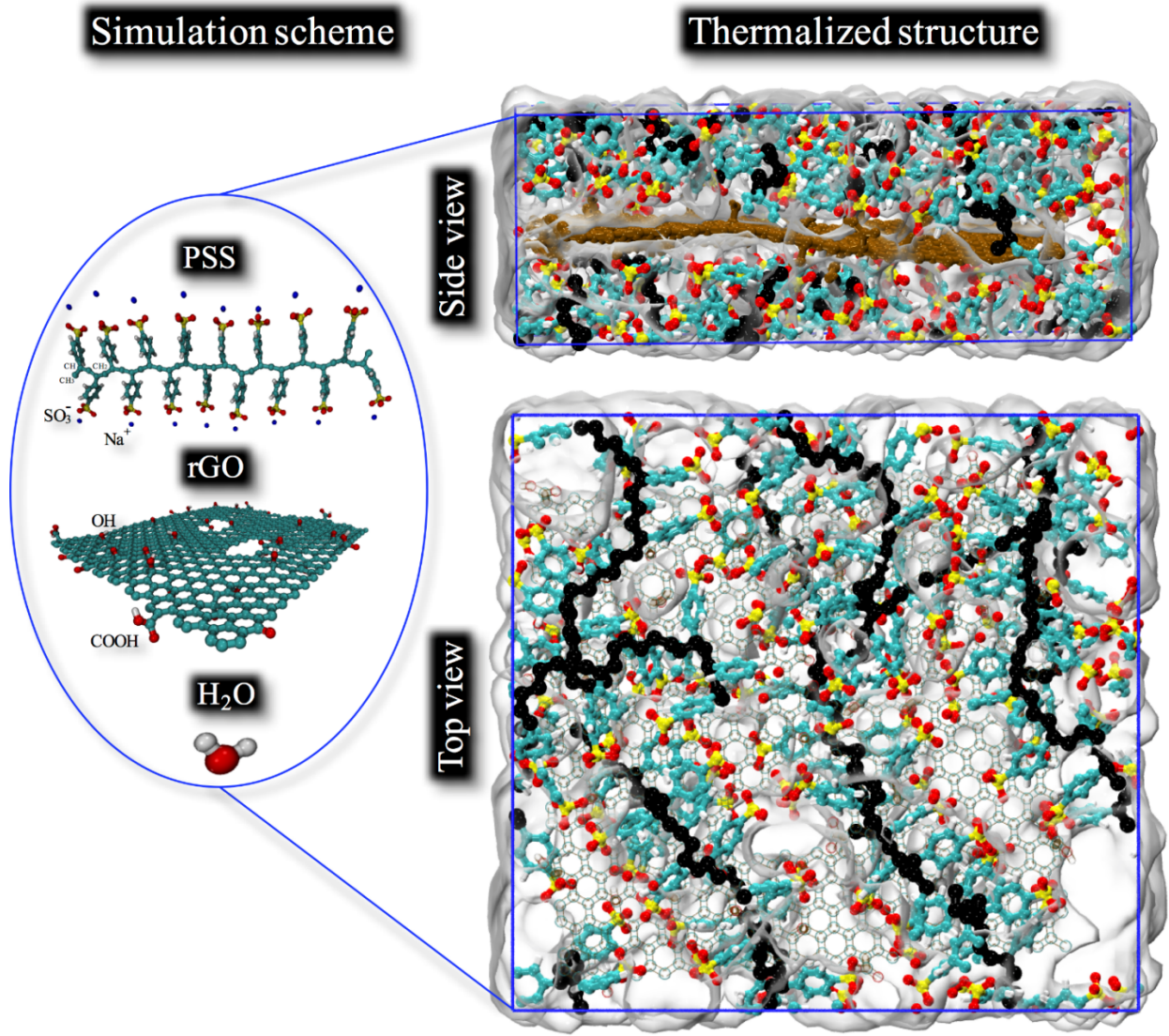

**Figure 1.** PPS and water molecules surrounding one single rGO nanoplatelet.

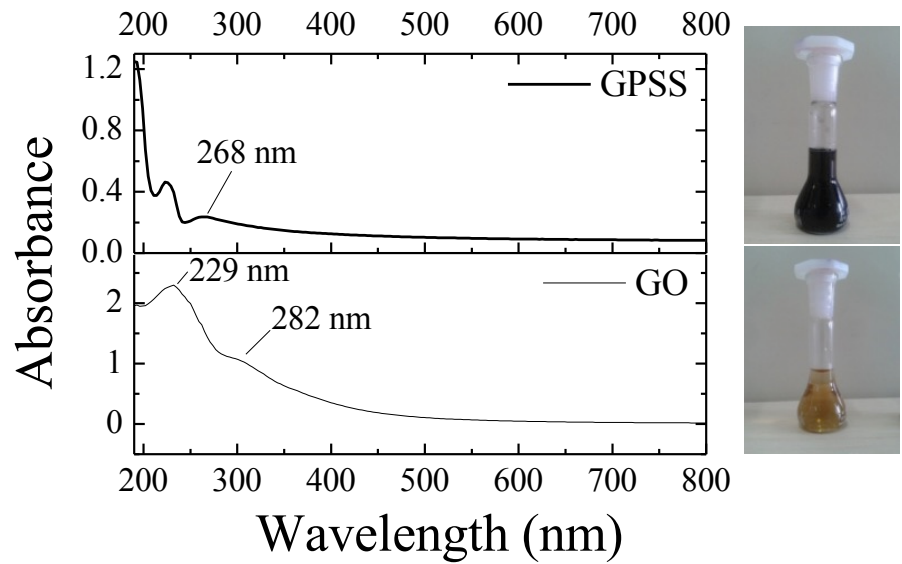

**Figure 2.** UV-vis absorption spectra of 0.1 g.L$^{-1}$ GO and GPSS dispersions.

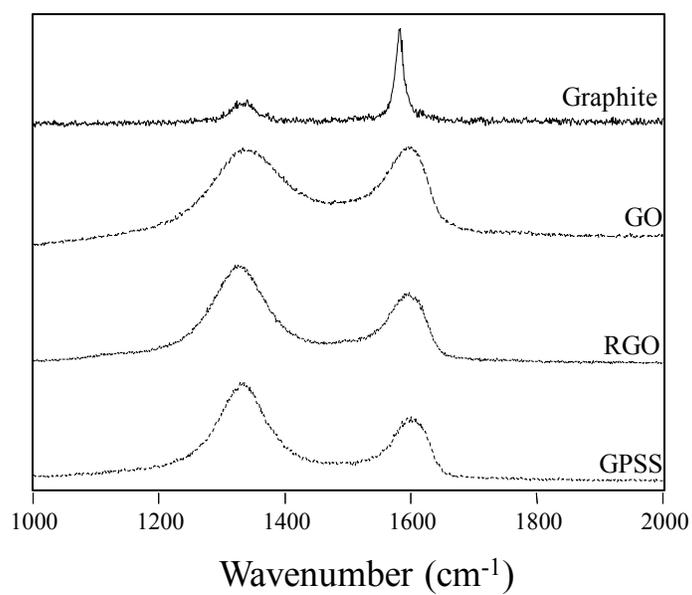

**Figure 3.** Raman spectra of graphite, GO, rGO and GPSS powder samples.

**Table 1.** Raman spectroscopy data for graphite, GO, rGO and GPSS powder samples.

|  | D peak position | G peak position | $I_D/I_G$ |
|---|---|---|---|
| Graphite | 1335 | 1582 | 0.25 |
| GO | 1336 | 1596 | 0.98 |
| rGO | 1325 | 1596 | 1.36 |
| GPSS | 1331 | 1599 | 1.53 |

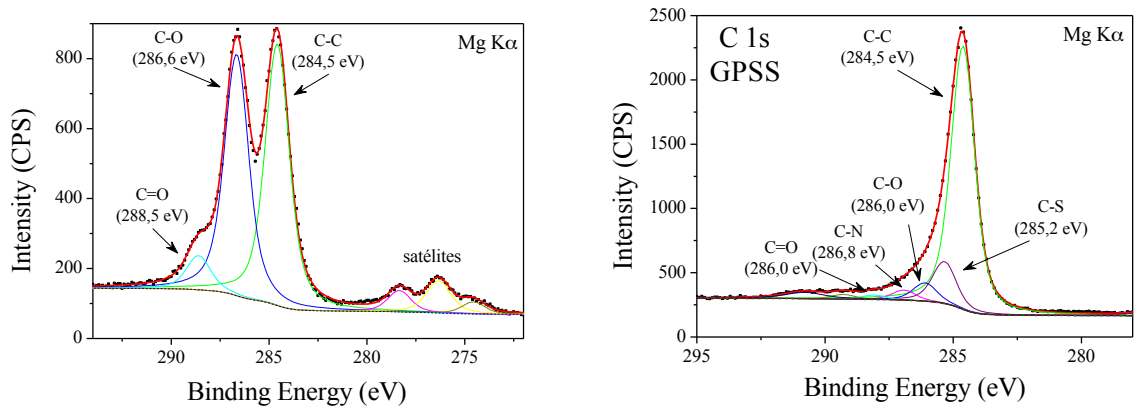

**Figure 4.** XPS spectra for GO and GPSS in the region of C1s.

**Table 2.** Relative percentage of functional groups compound the GO and GPSS samples.

| Sample | Functional group | % | Sample | Functional group | %* |
|---|---|---|---|---|---|
|  | C=C (rede sp$^2$) | 48.7 |  | C=C (sp$^2$ net) | 74.3 |
|  | C–O | 44.9 |  | C–S | 11.8 |
| GO | C=O | 6.5 | GPSS | C–O | 5.2 |
|  |  |  |  | O–C=O | 1.3 |
|  |  |  |  | C=O | 1.2 |

* GPSS spectrum presented 2.8 % signal from C in impurities and 3.4% from π- π * satellite.

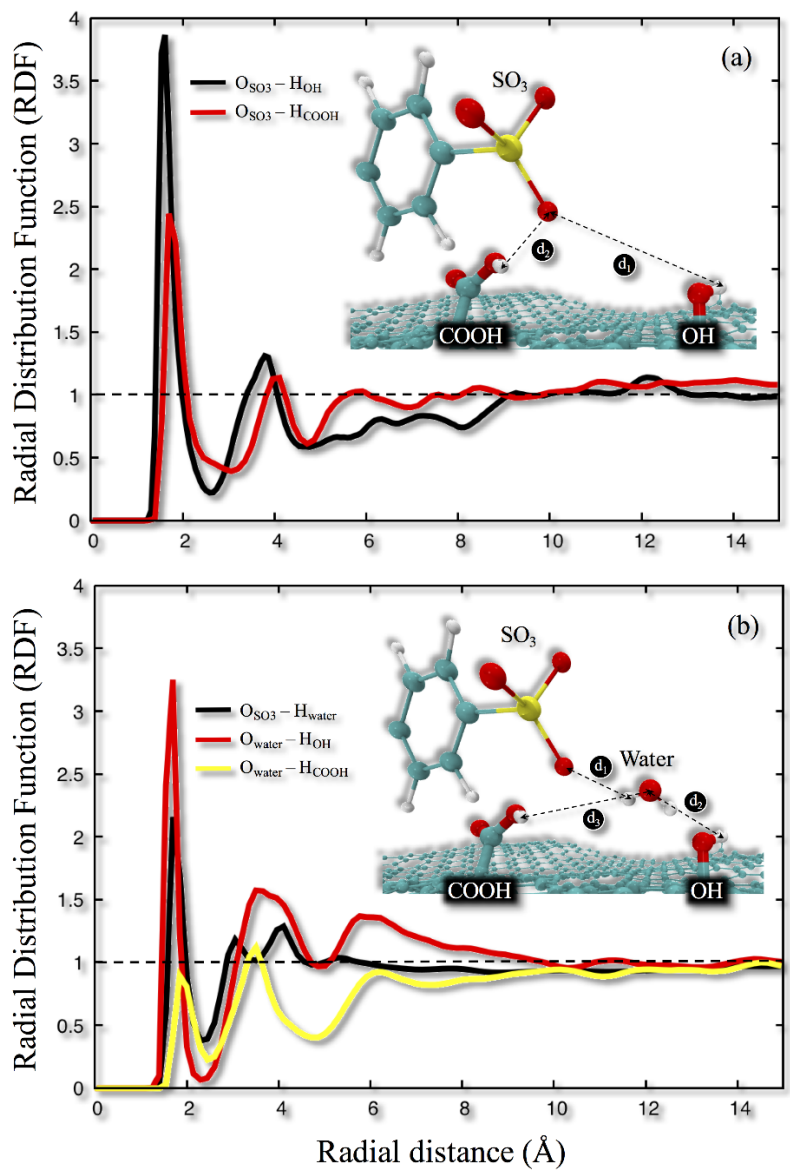

**Figure 5.** Radial pair distribution function, $g(r)$, of oxygen atoms in the PSS molecule and hydrogen atoms in carboxylic groups of the rGO nanoplatelet.

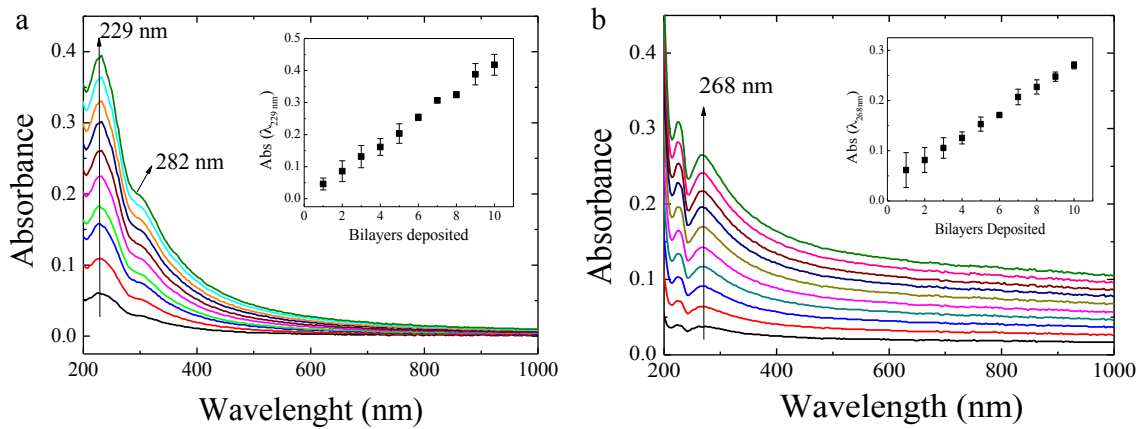

**Figure 6.** UV-vis absorption spectra of the LbL films growth onto quartz plates: (a) (PAH/GO)$_{10}$; (b) (PAH/GPSS)$_{10}$.

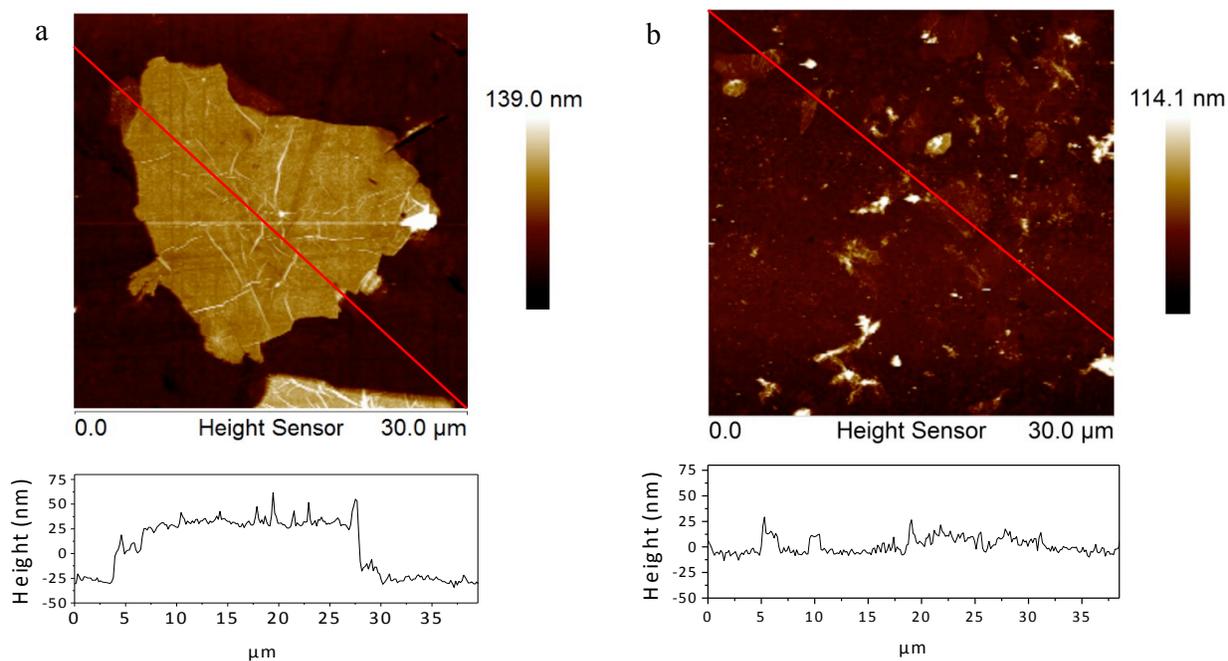

**Figure 7.** Tapping mode AFM images and height profile of (PAH/GO)$_5$ and (a) (PAH/GPSS)$_5$ (b) deposited onto quartz substrate.

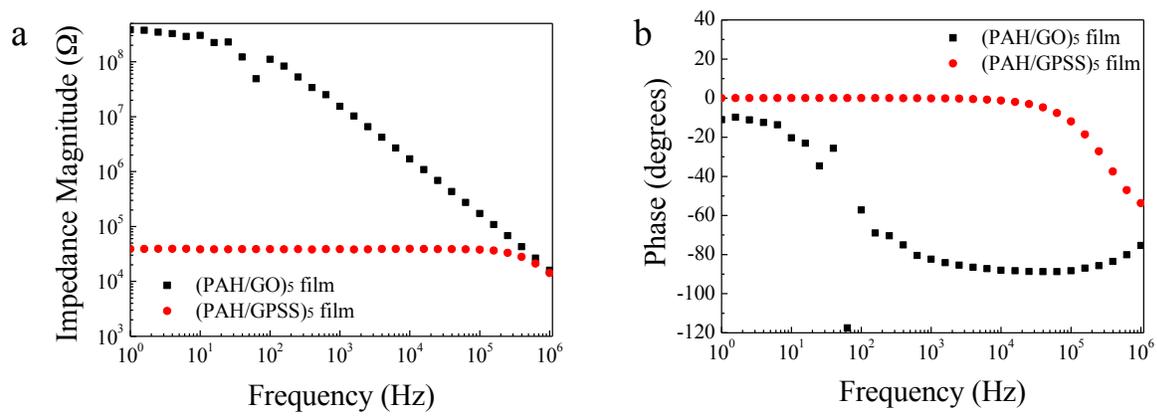

**Figure 8.** (a) Impedance modulus and (b) phase angle spectra of (PAH/GO)$_5$ and (PAH/GPSS)$_5$ LbL films.

**Acknowledgment**

Authors are grateful to FAPESP (2010/13033-6, 2012/01484-9, 2015/14703-9, 2016/00023-9 and 2016/12340-9), INEO (CNPq) and NanoBio Net (Brazil) for financial support, and also Bernhard Gross Polymer Group (IFSC – USP), Ângelo L. Gobbi and Maria Helena O. Piazzetta at Laboratory of Microfabrication (LNNano/CNPEM) for the cooperation in data acquisition and microelectrodes fabrication.